\documentclass[runningheads,fleqn]{svmult}

\usepackage{makeidx}   
\usepackage{graphicx}  
\usepackage{subeqnar}  
\usepackage{multicol}  
\usepackage{physproc}  
\makeindex             

\begin{document}
\title*{A Note on Stable States of Dipolar Systems\protect\newline at Low Temperatures}
\toctitle{A Note on Stable States of Dipolar Systems at Low Temperatures}
%
%
\titlerunning{A Note on Stable States of Dipolar Systems at Low Temperatures}
%
\author{Yusuke Tomita\inst{1}
\and Katsuyoshi Matsushita\inst{1}
\and Akiyoshi Kuroda\inst{1}
\and Ryoko Sugano\inst{2}
\and Hajime Takayama\inst{1}}
\authorrunning{Y. Tomita et al.}
%
%
\institute{Institute for Solid State Physics, University of Tokyo, Kashiwa, Chiba 277-8581, Japan
\and Advanced Research Laboratory, Hitachi, Ltd., Hatoyama-machi, Saitama 350-0395, Japan
}

\maketitle


\section{Introduction}
In the past several years, many important innovations in nanotechnology were made.
Today it becomes possible to make nanosize magnetic particles, 
and development of high storage-density magnetic device is desired.
Though dipole interaction plays the main role in these magnetic particle systems,
there is little systematic study of dipolar systems~\cite{Romano}.

\begin{figure}
\includegraphics[width=0.85\textwidth]{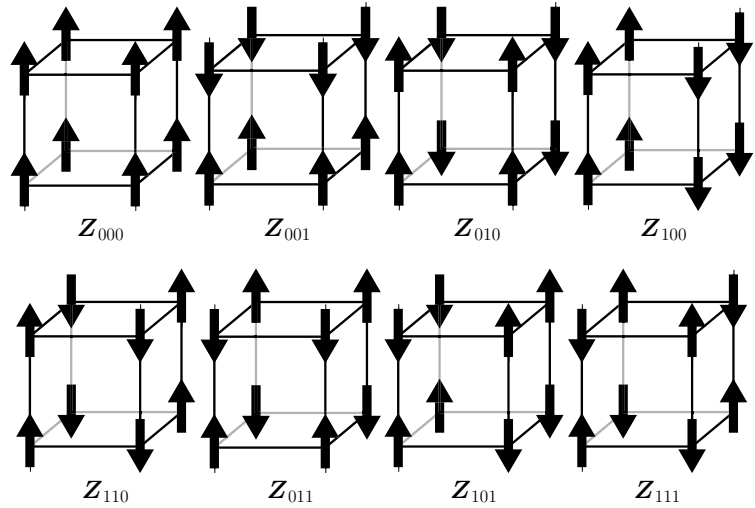}
\caption[]{The eight basic arrays.}
\label{BasicArrays}
\end{figure}

Luttinger and Tisza (LT) discussed the ground states of
three-dimensional dipole cubic lattices in their pioneering work~\cite{LuttingerTisza}.
They assumed $\vec{\Gamma}^2$ symmetry that ground states are constructed by
the translations of the dipoles on $2 \times 2 \times 2$ cubic unit,
and examined its dipole configurations.
In addition, since the dipole interaction energy is a quadratic form, 
even though dipole moment has $O(3)$ symmetry, it is sufficient to consider
only eight arrays for each component of the moment.
We define eight arrays by
\begin{equation}
\vec{A}_{b_x b_y b_z}(\vec{l}) = (-1)^{b_x l_x + b_y l_y + b_z l_z},
\qquad(b_x, b_y, b_z, l_x, l_y, l_z = 0, 1).
\end{equation}
Here $b_x, b_y, b_z$ are indices for a dipole configuration,
and an array $(l_x, l_y, l_z)$ represents a corner of the unit cube.
An eight-dimensional vector $\vec{A}$ corresponds to $\vec{X}$, $\vec{Y}$, or
$\vec{Z}$ depending on its component of moments.
In Fig.~\ref{BasicArrays}, we show the eight basic arrays.
%
%
For the simple cubic (SC) lattice, Luttinger and Tisza predicted the columnar
antiferromagnetic state,
\begin{equation}
a\vec{X}_{011} + b\vec{Y}_{101} + c\vec{Z}_{110},\qquad(a^2 + b^2 + c^2 = 1),
\label{GS_SC}
\end{equation}
has the lowest energy. For the body centered cubic (BCC) lattice, lowest energy
states predicted by LT are tabulated in Table~\ref{TableBCC}.
%
\begin{table}
\caption{Lowest energy states are tabulated below.
The states of lattice points (l.p.) and body centers (b.c.) are commutative.
The condition, $a^2 + b^2 + c^2 = 1$, is satisfied for the
first line, and others satisfy the condition,
$a^2 + b^2 = 1$.}
\begin{tabular}{cc}
\hline
At l.p.(b.c.) & At b.c.(l.p.)\\ \hline
$a\vec{X}_{101} + b\vec{Y}_{110} + c\vec{Z}_{011}$ &
$b\vec{X}_{110} + c\vec{Y}_{011} + a\vec{Z}_{101}$\\
$a\vec{X}_{101} + b\vec{Z}_{101}$ & $b\vec{X}_{101} + a\vec{Z}_{101}$\\
$a\vec{Y}_{110} + b\vec{X}_{110}$ & $b\vec{Y}_{110} + a\vec{X}_{110}$\\
$a\vec{Z}_{011} + b\vec{Y}_{011}$ & $b\vec{Z}_{011} + a\vec{Y}_{011}$\\
\hline
\end{tabular}
\label{TableBCC}
\end{table}
%

\section{Results}
\begin{figure}
\includegraphics[width=0.75\textwidth]{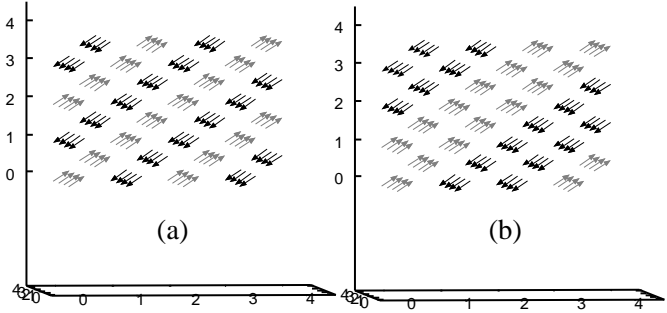}
\caption[]{Stable states for the body centered cubic lattice. 
(\textbf{a}) A stable state which is consistent with LT's predicted ground states.
(\textbf{b}) A stable state obtained by MC simulation which is not consistent with
the states in Table~\ref{TableBCC}.}
\label{BCC4}
\end{figure}

In order to examine the lowest energy states predicted by LT,
we simulate dipolar systems on the SC lattice and the BCC lattice
at low temperatures. The Hamiltonian for system size $L$ is given by
\begin{equation}
{\cal H} = \sum_{n_x, n_y, n_z}
\sum_{i < j}\left[\frac{\vec{\mu}_i \cdot \vec{\mu}_j}{(\vec{r}_{ij}+\vec{n} L)^3}
- 3\frac{(\vec{\mu}_i \cdot (\vec{r}_{ij}+\vec{n} L))(\vec{\mu}_j \cdot (\vec{r}_{ij}+\vec{n} L))}
{(\vec{r}_{ij}+\vec{n} L)^5}\right].
\end{equation}
We employed heat bath method for Monte Carlo spin update, 
and Ewald summation method~\cite{Ewald,Kornfeld,LeeuvEtAl}
is used for counting long range dipolar interaction.
As expected, for the SC and BCC of $L=2$ lattice systems,
dipole configurations produced by simulations well agree with the prediction by LT.
The dipole configurations are consistent with the lowest energy state
described in (\ref{GS_SC}) for the SC of $L \ge 4$ lattices.
On the other hand, for the BCC of $L \ge 4$ lattices,
the stable dipole configuration that we obtained are different from
the states in Table~\ref{TableBCC}, which is shown Fig.~\ref{BCC4}(\textbf{b}).
%
%
The dipole configuration in Fig.~\ref{BCC4}(\textbf{b}) is characterized by
stacked dipole layers in which all dipoles align in the same direction.
Such stacked dipole layer configurations can be constructed from LT's predicted state,
although ferromagnetic layer is single (Fig.~\ref{BCC4}(\textbf{a})),
but not double as in Fig.~\ref{BCC4}(\textbf{b}).
We calculate the zero temperature internal energy as a function of thickness of ferromagnetic
layer to check our result of MC simulation.
The results are shown in Table~\ref{TableEng}.
At least till $L=8$, the internal energy becomes lower as the thickness of ferromagnetic
layer increases.
Furthermore the values of these internal energies are lower than that of LT calculated.
This means that one cannot assume $\vec{\Gamma}^2$ symmetry to dipole systems
on the BCC lattice.
%
\begin{table}[t]
\caption{The zero temperature internal energy dependence on thickness of ferromagnetic layer.}
\begin{tabular}{ccccc}
\hline
Thickness of FM layer & $L=2$ & $L=4$ & $L=6$ & $L=8$\\ \hline
1 & -7.94368 & -7.94368 & -7.94368 & -7.94368\\
2 & --- & -8.16968 & --- & -8.16968\\
3 & --- & --- & -8.23896 & ---\\
4 & --- & --- & --- & -8.27087\\
\hline
\end{tabular}
\label{TableEng}
\end{table}
%

We set up a simplified model to obtain an intuitive understanding of stable states
of dipolar systems at low temperatures.
In some cases, consideration of low dimensional system provides insights into
physics of the system. So we start looking at two-dimensional systems.

\begin{figure}[t]
\includegraphics[width=0.75\textwidth]{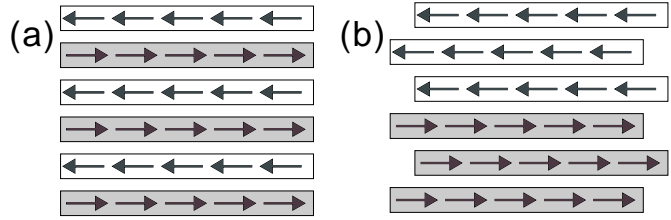}
\caption[]{(\textbf{a}) A typical dipole configuration in the ground state on the square lattice.
(\textbf{b}) A typical dipole configuration in the ground state on the triangular lattice.}
\label{dim2conf}
\end{figure}

Typical ground states of the square lattice and triangular systems are depicted
in Fig.~\ref{dim2conf}.
In order to investigate the stable state at low temperatures, we simplify the model as follows:
we treat the system as interacting ferromagnetic dipole chain system, and ferromagnetic chain
can take only two values (rightward or leftward as in Fig.~\ref{dim2conf}).
Then, Hamiltonian becomes
\begin{equation}
{\cal H} = \sum_{i \ne j}f(r_{ij})\sigma_i\sigma_j, \qquad(\sigma = \pm 1),
\end{equation}
that is, the model is mapped to a one-dimensional Ising chain with long-range
interaction $f(r)$.
We estimate $f(r)$ by numerical calculation with $L/2$ cut off.
At the zero temperature internal energy of the square lattice and the triangular lattice was
calculated for several ferromagnetic dipole chain configurations.
For the square lattice, antiferromagnetic state is the most stable, whereas
for the triangular lattice, antiparallel two ferromagnetic domains
structure is the most stable.
This difference comes from the shape of $f(r)$, that is, $f(r)$ for square lattice
is positive and monotonically decreasing function.
For the triangular lattice, on the other hand, except for nearest neighbor, 
$f(r)$ is positive and its absolute value decreases monotonically.
That is to say, ordering in the same direction gains energy in short range,
while ordering in opposite direction gains energy in long range; as a
consequence, ferromagnetic domains structure is realized at low temperatures.

\begin{figure}[t]
\includegraphics[width=0.75\textwidth]{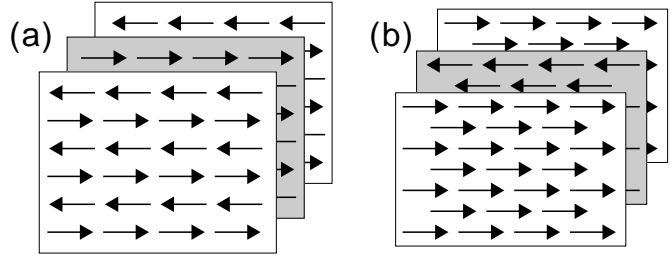}
\caption[]{(\textbf{a}) A typical dipole configuration in the ground state on the SC lattice.
(\textbf{b}) A typical dipole configuration in the ground state on the BCC lattice.}
\label{dim3conf}
\end{figure}

\begin{figure}
\includegraphics[width=0.80\textwidth]{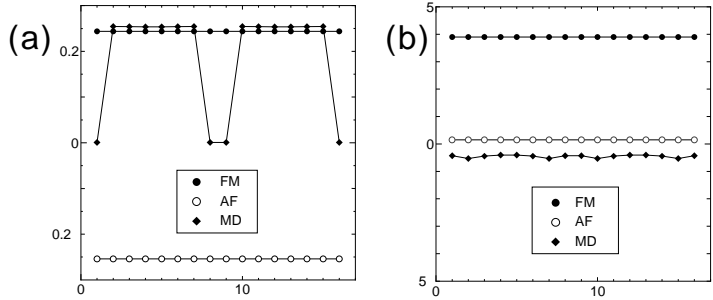}
\caption[]{(\textbf{a}) The internal energy at each site (layer) of the SC lattice ($L=16$).
The internal energy per site is $E_{\rm FM}=0.2437$, $E_{\rm AF}=-0.2437$,
and $E_{\rm MD}=0.1907$ for ferromagnetic dipole configuration, antiferromagnetic
dipole configuration, and magnetic domain dipole configuration respectively.
(\textbf{b}) The internal energy at each site (layer) of the BCC lattice ($L=16$).
The internal energy per site is $E_{\rm FM}=3.9016$, $E_{\rm AF}=0.1488$,
and $E_{\rm MD}=-0.4508$ for each dipole configuration.}
\label{dim3eng}
\end{figure}

Since the simplified model succeeded in explaining dipole configurations on
two-dimensional lattices at low temperatures, we extend the model to three-dimensional systems.
In Fig.~\ref{dim3conf}, we show typical ground states for the SC lattice and
the BCC lattice. 
%
%
%
%
If one notes dipole configuration on the gray sheet
is merely reversed configuration on the white sheet, it is natural to apply the same
tactics which is employed in the two-dimensional system to the three-dimensional
systems: there are magnetic sheets interacting each other, and a magnetic sheet
can take only two values as like Ising variable. Then, we obtain simplified Hamiltonian,
${\cal H} = \sum_{i \ne j}f(r_{ij})\sigma_i\sigma_j.$
Again, the model is mapped to a one-dimensional Ising chain with long-range
interaction. In Fig.~\ref{dim3eng}, we show the results of the SC lattice and
the BCC lattice. 
%
%
The result of the SC lattice supports the one of LT.
On the other hand, the result of the BCC lattice
is not consistent with the LT result, but
with our results of MC simulation. It is notable that ferromagnetic domains
structure is stable for the BCC lattice, even though interaction
$f(r)$ is positive.
We also examined $L=64$ chain model, and we confirmed the four ferromagnetic
domains structure is the most stable.

\section{Summary}
To summarize, we showed ``antiferromagnetic structure'' is stable for the square lattice
and the SC lattice as LT predicted. On the other hand, for the triangular lattice
and the BCC lattice, magnetic domain structure is stable. Theoretical approach which
assumes $\vec{\Gamma}^2$ symmetry fails in the triangular lattice
and the BCC lattice.
For future works, examination of the face centered cubic lattice and estimation of
critical temperatures and exponents for several lattices are remained.

\end{document}